\documentclass{PoS}
\usepackage{graphicx}
\usepackage[subfigure]{graphfig}
\usepackage{subfig}
\usepackage{amsmath}
\usepackage{epsfig}
\usepackage{epstopdf}

\def\be{\begin{equation}}
\def\ee{\end{equation}}
\def\bea{\begin{eqnarray}}
\def\eea{\end{eqnarray}}

\definecolor{green}{rgb}{0,.5,0}

\title{Nucleon Matrix Elements at Physical Pion Mass and Cost Comparison
\thanks{This work is supported in part by the U.S. DOE Grant No. DE-SC0013065.
}}

\ShortTitle{Nucleon Matrix Elements}

\author{\speaker{Keh-Fei Liu}, Jian Liang and Yi-Bo Yang
\vspace*{-0.5cm}
\begin{center}
\large{
\vspace*{0.4cm}
\includegraphics[scale=0.20]{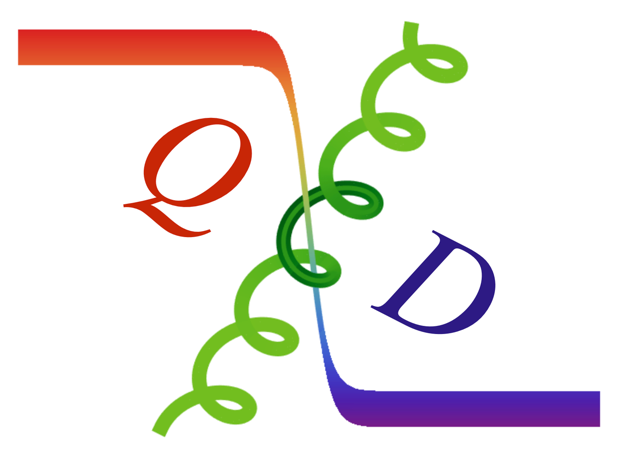}\\
\vspace*{0.4cm}
($\chi$QCD Collaboration)
}
\end{center}
\\
\mbox{Department of Physics and Astronomy, University of Kentucky, Lexington, KY 40506, USA}\\
}

\abstract{\hspace{0.8cm}
We report a lattice calculation of nucleon forward matrix elements  on a  
$48^3 \times 96$ lattice at the physical pion mass and a spatial size of 5.5 fm. The $2+1$ flavor dynamical fermion configurations are generated with domain-wall fermions (DWF) and the overlap fermions are adopted for the valence quarks.  The isovector $g_A^3$ and  $g_S^3$, and the connected insertion part of $g_S^0$ are reported for three source-sink separations. With local current, we obtain
$g_A^3 = 1.18(4)$ from a two-state fit. For the quark momentum fraction $\langle x \rangle_{u-d}$, we have included
smaller lattices (i.e. $24^3 \times 64$ and $32^3 \times 64$ lattice with pion mass at 330 and 290 MeV respectively) for a 
fit which includes partially quenched cases as well as finite volume and continuum corrections. A global fit with perturbative 
renormalization gives $\langle x \rangle_{u-d} (\overline{MS},\, \mu = 2\, {\rm GeV}) = 0.170(14)$. 

\hspace{0.8cm}
We made a cost comparison of calculating the nucleon matrix elements with those from the twisted mass fermion on similar sized lattice at the physical pion point and the domain-wall fermion calculation on the same DWF lattice. We also compare cost with the clover fermion calculation on similar sized lattice at about the same quark mass. The comparison shows that with several improvements, such as many-to-all correlator with grid source and low-mode substitution in the connected insertion and low-mode average in the quark loop can make the overlap as efficient as the twisted-mass and clover fermions in calculating the three-point functions.  It is more efficient than the DWF. When the multi-mass feature is invoked, the overlap can be more efficient in reaching the same precision than the single
mass comparison made so far.}

\FullConference{34th International Symposium on Lattice Field Theory - LATTICE 2016\\
        July 24 - July 30, 2016\\
        Highfield Campus, University of Southampton, UK}

\begin{document}

       With the advent of chiral fermions such as the domain-wall fermion and overlap fermion which observes chiral symmetry
{\it \'{a} la} Ginsparg-Wilson relation, an issue one faces is the following. Should one proceed with the chiral fermion which
is one or two orders of magnitude more expensive than the other non-Ginsparg-Wilson fermion actions (i.e. staggered, clover, twisted-mass fermions, etc.) to invert the fermion matrix in order to calculate physical observables, some of which are sensitive to spontaneously broken chiral symmetry at finite lattice spacing and hope that the $\mathcal{O}(a^2)$ errors are small so that one can carry out the continuum limit extrapolation from relatively coarser lattice spacings? Or should one proceed with the other fermions which are much cheaper numerically and perform the calculation closer to the continuum and hope that chiral symmetry is recovered so that one can obtain a reliable result faster? It is, in some sense, a choice between expediency and principle. In the end, which path is more desirable depends on whether the correct results with all the systematics (physical pion mass, continuum and infinite volume limits, chiral symmetry) taken into account are obtained with less resources. 

       Since there is an unduly burden to demonstrate that the continuum limit is reached with proper spontaneously broken chiral symmetry for the non-Ginsparg-Wilson fermions, we have chosen to use the overlap fermion, which satisfies the Ginsparg-Wilson relation to machine precision and thus should have a smooth approach to the continuum with the correct chiral symmetry. In this case, it contingent upon the practitioners to show that correlators can be constructed efficiently to make up for the longer inversion time as compared to the other fermion formulations. 
       
       In the past several years, we have developed several algorithms and techniques, some of which are based on the
salient features of the overlap fermion,  to enhance statistics of the correlators. First of all,
the eigenmode deflation is adopted to speed up the inversion of the overlap fermion for both the Wilson fermion kernel $H_w$ and 
the overlap operator $D_{ov}$~\cite{Li:2010pw}. The low-energy eigenmodes are then used to construct many-to-all correlators 
to replace the noise-estimated low-frequency part of the hadron correlators with the exact one.  
It is found~\cite{Gong:2013vja} that the smeared-grid noise source with low-mode
substitution (LMS)  gains in statistics about $\sim 60\%$ of  the number of grids (8 for smaller 
lattices and 64 for the $48^3 \times 96$ lattice) as compared to a single smeared noise 
source without LMS for the nucleon two-point function. This, together with the use of the low-mode substitution (LMS) technique described in Ref.~\cite{Yang:2015zja,Gong:2013vja}, allows us to obtain hundreds of measurements with just a few inversions, 
thus overcoming the expensive cost of the overlap action required to obtain more precise results.
We have also calculated quark loops for the scalar and pseudoscalar
densities in the nucleon with low-mode average (LMA) and noise estimate of the high 
modes~\cite{Gong:2013vja,Gong:2015iir,Yang:2015uis}.  Due
to the dominating low-mode contribution in the scalar and pseudoscalar loops as observed, we
have been able to calculate, with the overlap fermion on DWF configurations,
the strangeness and charmness contents in the nucleon with very high precision amongst
all the current lattice calculations and at a smaller fraction of the cost compared to other 
calculations~\cite{Gong:2013vja}.  Our  most recent work on the strangeness
content~\cite{Yang:2015uis} has a similarly high precision among all the $2+1$-flavor lattice 
calculations and it has included the physical pion point from the 48I lattice as well as finite volume 
and continuum corrections. The speed-up in inversion and
the improvement in correlator and quark loop calculation are testaments to the
fact that low eigenmodes are crucial to all the above improvements.

Besides the LMS for the nucleon propagator and LMA for the quark loop, both of which
are important for disconnected insertion calculations, we have implemented a stochastic
sandwich method for the three-point connected insertion (CI) calculations~\cite{Yang:2015zja}.
This involves a multi-grid smeared source with LMS and stochastic sink for the high modes. 
This saves time as compared to the usual sink sequential method, the latter 
needs to have multiple inversions at the sink for both $u$ and $d$, different polarizations, and different
momentum. Our improved stochastic method replaces the long-distance part of the stochastic propagator 
from the sink to the current by its all-to-all version, using the low-lying eigensystem of $D_{ov}$ , which 
suppresses the influence of the stochastic noise on the sink propagator.
 In addition, the stochastic sandwich method can accommodate muti-mass
inversions which can further reduce errors in global fittings with  partially quenched data, even though they
are correlated.


In this proceedings, we use the valence overlap fermions on $2 +1$-flavor domain-wall fermion (DWF) 
configurations~\cite{Blum:2014tka} at the physical pion mass to carry out the calculation. The relevant parameters of the lattice
are listed in Tabel~\ref{table:r0}.


\begin{table}[htbp]
\begin{center}
\caption{\label{table:r0} The parameters for the RBC/UKQCD configurations\cite{Blum:2014tka}: spatial/temporal size, lattice spacing, the sea strange quark mass under $\overline{MS}$ scheme at {2 GeV}, the pion mass with the degenerate light sea quark, and the number of configurations ($N_{cfg} $) used in this work.}
\begin{tabular}{ccccccc}
Label & $L^3\times T$  &a (fm)  &$m_s^{(s)}$ (MeV) &  {$m_{\pi}$} (MeV)   & $N_{cfg} $ \\
\hline
48I &$48^3\times 96$& 0.1141(2) & 94.9   &139 & 81  \\
\hline
\end{tabular}
\end{center}
\end{table}


 A regular grid with 4 smeared $Z_3$-noise sources in each spatial direction for the 48I lattices are placed
on 3 time slices. The separation between the centers of the neighboring grids is $\sim 1.3$ fm and each smeared source has a radius of $\sim$ 0.5 fm.
On the sink side, several noise point-grid sources are placed at three slices $t_f$ which are $0.9-1.4$ fm away from the source time slices so that the source-sink time separation is $t_{sep} = t_f - t_0$. Furthermore, the matrix elements of the light scalar contents are dominated by the low-mode part of $D_c$ so that the use of LMS on the propagators from the current to the sink notably reduces the number of noise propagators (from $t_f$) needed~\cite{Yang:2015zja}.

 The same noise grid-smeared sources are used in the production of the nucleon propagator for the disconnected insertion, and we loop over all the time slices for the nucleon source. The position of the grid is randomly shifted on each time slice.
As has been carried out
in previous studies of the strangeness content~\cite{Gong:2013vja} and quark spin~\cite{Gong:2015iir}, the quark loop is calculated 
with the exact low eigenmodes (low-mode average (LMA)) while the high modes are estimated with 8 sets of $Z_4$ noise on the same (4,4,4,2) grid with odd-even dilution and additional dilution in time. 

 \begin{table}[htbp]
\begin{center}
\caption{\label{table:compare}
Comparison of the nucleon matrix elements calculated with the overlap fermion and the twisted-mass fermion with clover term (TM+C)
~\cite{Abdel-Rehim:2015owa} for three sink-time separations. } 
\begin{tabular}{|c|ccc|ccc|}
\hline
    \multicolumn{4}{|c}{\hspace{2cm}Overlap} & \multicolumn{3}{c|} {TM+C} \\
\hline
\hline
 $t_{sep}$   &  0.91 fm &  1.14 fm&  1.37 fm &0.90 fm & 1.08 fm & 1.26 fm\\
 \hline
$g_A^3$ & 1.133(15)&1.150(25)&1.233(66)&1.158(16)&1.162(30)&1.242(57)\\
$g_S^3$ & 0.72(8) & 0.93(17) & 0.78(41) & 0.55(18)  &1.18(34) & 2.20(54)\\
$g_S^0$(CI) & 6.80(15) &7.23(33) &7.77(70) &6.46(27) &7.84(48)& 8.93(86)    \\
$\langle x \rangle_{u-d}$ & 0.214(9)&0.194(11)&0.195(28)&0.248(9)&0.218(15)&0.208(24)\\
$\langle x \rangle_{u+d}$(CI) & 0.519(11)&0.456(15)&0.400(36)&0.645(13)&0.587(18)&0.555(63)\\
\hline
\end{tabular}
\end{center}
\end{table}  

     In Table~\ref{table:compare}, we compare our results on the 48I lattice with those from the twisted-mass fermion with clover term 
(TM+C)~\cite{Abdel-Rehim:2015owa}. Both calculations used ensembles with the size $48^3\times 96$. TM+C has a lattice spacing of $0.093$ fm and pion mass of 131 MeV. Both are close to those of the overlap 48I lattice (see Table~\ref{table:r0}). 
Listed in Table.~\ref{table:compare} are the isovector $g_A^3$, $g_S^3$ and the $u$-$d$ quark momentum fraction difference 
$\langle x \rangle_{u-d}$. Also listed are the connected insertion part of isoscalar $g_S^0$ (CI) and $\langle x \rangle_{u+d}$ (CI) for three source-sink separations. $g_A^3$ is normalized with chiral Ward identity, $g_S^{3,0}$ are non-perturbatively renormalized
with the RI-MOM scheme and matched to $\overline{MS}$ at $\mu = 2$ GeV. The momentum fraction $\langle x \rangle_{u\pm d}$
are perturbatively renormalized and matched to $\overline{MS}$ at $\mu = 2$ GeV. 
We see that, except for a few cases, they are, by and large, in agreement within two-sigma errors.
Actually, the error bars of the overlap results are slightly smaller that those of the TM+C for the two smaller $t_{sep}$ which can be
compared directly, since they are quite close to each other in these two calculations. We shall compare the cost of these two calculations to reach these comparable errors. Before we do that, we would like to discuss the results of  $g_A^3$ and $\langle x \rangle_{u-d}$.

 \begin{figure}[hbt]
 \centering
 \subfigure[$g_A^3$]{ \includegraphics[width=0.4\textwidth]{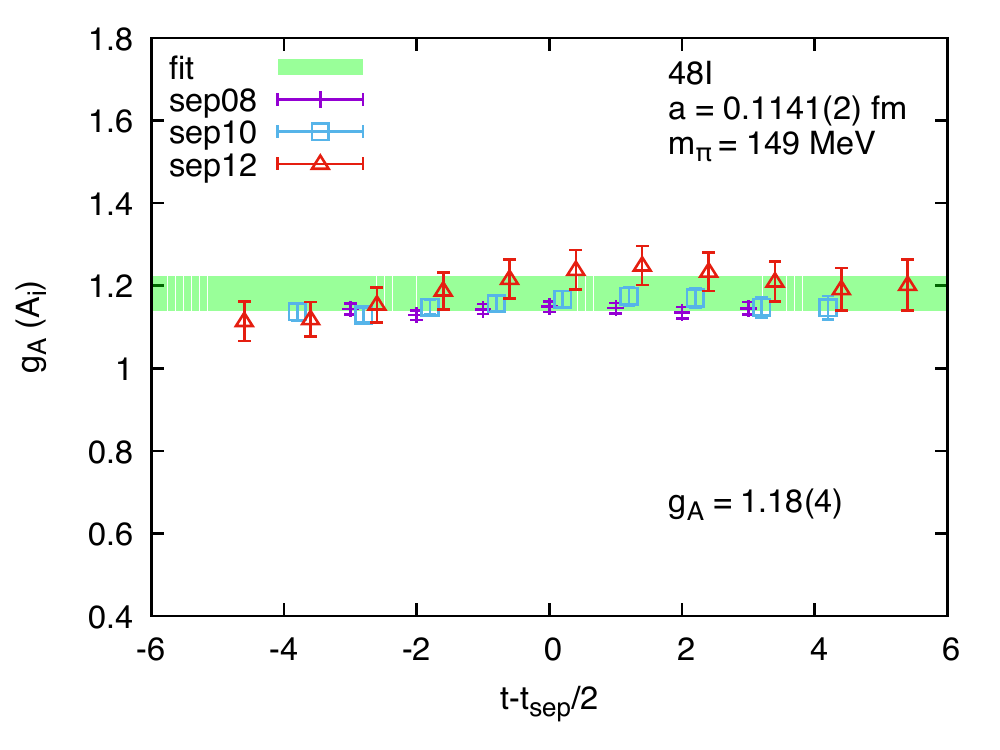}}
 \subfigure[$\langle x \rangle_{u-d}$]{ \includegraphics[height=0.31\hsize,width=0.48\textwidth]{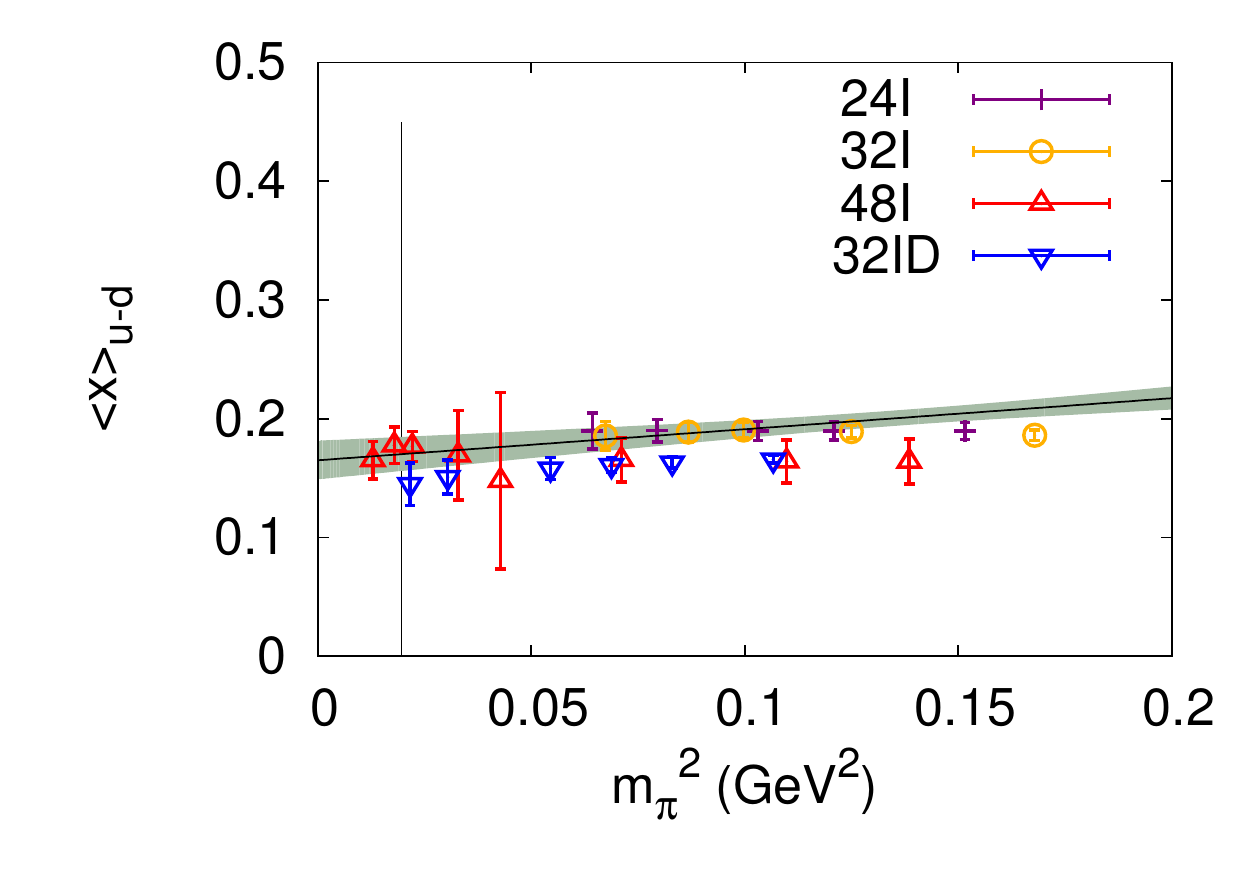}}
\caption{(a) Isovector $g_A^3$ as a function of the $t - t_{sep}/2$  where $t$ is the time position of the current insertion. Time separation $t_{sep}$ between the nucleon source and sink at 8, 10, and 12 lattice units are plotted. The band shows the two-state fit with these 3 separations. (b) The difference between the $u$ and $d$ quark momentum fractions $\langle x \rangle_{u-d}(\overline{MS}, 2 \,{\rm GeV})$  as a function of $m_{\pi}^2$ for 4 lattices. The band shows the global fit from these 4 lattices and the vertical line indicate the position of the physical pion mass. 
\label{gAx} } 
\end{figure}

Plotted in Fig.~\ref{gAx}(a) are the ratios of 3-pt to 2-pt functions as a function of $t - t/_{sep}2$ to obtain $g_A^3$, where $t$ is the position of the current insertion relative to the source and and time separation $t_{sep}$ at 8, 10, and 12 lattice units between the nucleon source and sink are plotted. 
Using two-state fit, we obtain $g_A^3 = 1.18(4)$ which is still somewhat lower than the experimental value of 1.2723(23). It is shown recently that the local axial-vector current suffers from an $\mathcal{O}(a)$ error and including an improvement operator raises
the $g_A^3$ value by 3.4\% on the smaller lattice $24^3 \times 64$ with same lattice spacing and a pion mass of 
330 MeV~\cite{Liang:2016fgy}. Assuming the same increase applies to the 48I lattice, we obtain $g_A^3 = 1.22(4)$ which is
about one sigma from the experimental value. Future calculation with conserved current and larger time separation are needed to settle
this potential discrepancy which has so far plagued many lattice calculations. 

We have performed a two state fit of the perturbatively renormalized $\langle x \rangle_{u - d}$ on the 48I lattice with several
valence masses as well as on several small lattices -- $24^3 \times 64$, $32^3 \times 64$, and $32^3 \times 64$ (DSDR) lattices at
three different lattice spacings and with sea pion masses of 330, 290, and 170 MeV, respectively. A global fit with the
formula $c_0 +  c_1\,m_{\pi,vv}^2 +  c_2\, m_{\pi,vs}^2 + c_3\, a^2 + c_4\, e^{-m_{\pi,vv}L}$, where $m_{\pi, vv}$ is the pion mass
from the valence quarks and $m_{\pi, vs}$ is the mixed valence-sea pion mass to incorporate continuum and finite volume corrections, and pion mass interpolation to the physical pion
mass gives \mbox{$\langle x \rangle_{u - d} = 0.170 (14)$.} This is quite a bit smaller than previous lattice calculations and is within one sigma from the NMC result of $\sim 0.161 (11)$~\cite{Ama91}.

To compare the numerical cost with the TM+C for both the CI~\cite{Abdel-Rehim:2015owa} and DI~\cite{Abdel-Rehim:2016won} calculations on the similar lattices and at physical pion mass, we list the relevant parameters for the two calculations in 
Table~\ref{twisted}. The overlap inversion for high precision is $\sim 10$ times slower than that of the twisted 
mass on GPUs of the same model (M2070)~\cite{KH15}. 

 \begin{center}
 \begin{table}[htbp]
  \centering
  \caption{Numbers of configurations, sources and total inversions are listed together with the relative inversion times and  errors for the calculation of the nucleon matrix elements in CI  and strangeness in DI for the TM+C and overlap fermions.   Loop is the number of equivalent high precision noise propagators for the loop calculation in DI. LMS denotes the overhead due to the low-mode substitution.}
{\small
  \begin{tabular} {|c|cccccc|}
  \hline
  Fermion  &  Config. & Source   & Inversion  & Inversion Time  &  LMS & Error   \\
    \hline
   TM+C (CI) & 96 & 16 & 16+16$\times 8\times$ 3 & 1  & 0 &  x\%  \\
    \hline
    Overlap (CI)   & 81 & 5 & 5+4+8+12  & $\sim 10$ & 1 inv/source & x\%  \\
    \hline\hline
   Fermion   &  Config. & Source   &  Loop & Inversion time  &  LMS & Error   \\ 
   \hline
   TM+C (DI) & 1800 & 100 & 14 &  1  & 0 &  20\%  \\
    \hline
    Overlap (DI)   & 81 & 32 &  32/3 & $\sim 10$ & 0 & 50\% (29\% multi-mass) \\
    \hline  
  
 \end{tabular} 
 }  \label{twisted}
 \end{table}
 \end{center}

 The cost ratio $R_{cost}$ of the other fermion to the overlap fermion to reach the same precision for the nucleon matrix elements is 
\begin{equation}  \label{Rcost}
R_{cost} = R_{\rm conf} \times R_{\rm inversion} \times R_{\rm time}  \times (R_{\rm error})^2
\end{equation}
where $R_{\rm conf}$ is the ratio of the number of configurations, $R_{\rm inversion}$ is the ratio of the number of
inversions which includes the overhead due to LMS, $R_{\rm time}$ is ratio of inversion time, and $R_{\rm error}$
is the ratio of errors. For the case of the CI, the number of inversion for TM+C sink sequential is the number of sources (16) times two inversions for $u$ and $d$, times four inversions for polarization and times the number of sinks (3) at different time slices.
On the other hand, the number of inversion for the overlap is the sum of the number of sources (5) and the numbers of
inversions at different sink time positions (4, 8, and 12) for the stochastic sandwich approach and it also includes the overhead for the
LMS which is 5 in this case.  Taking the errors to be the same in these two calculations, we obtain $R_{cost} = 1.4 $ for the CI calculations of the two approaches. Although the inversion of the overlap is slower by an
order of magnitude, the cost efficiency is imporved by using 64 grid sources on the source time slices with LMS and stochastic sandwich approach to save inversions at the sink. 

For the DI, we compare the strange quark sigma term $\sigma_{sN}=m_s \langle N| \bar{s}s|N|\rangle$ from
TM+C~\cite{Abdel-Rehim:2016won} and overlap~\cite{Yang:2015uis}. The total number of inversions in this case is
the sum of those for the nucleon propagators (Source) and the equivalent quark loop propagator from the noise with high precision. For the loop calculation with noise, we have divided the noise propagator inversion by a factor of 3 for the overlap to convert it to the equivalent high precision inversion. From Table~\ref{twisted}, we find $R_{cost} = 0.94$. When the global fit from partially quenched data with multi-masses is carried out, the error is decreased from 50\% to 29\% and, as a result, $R_{cost} = 2.8$. We see that the overlap with grid source and low-mode average (LMA) turns out to be about as efficient as the TM+C calculation, and it is more efficient when the multi-mass feature is utilized.

We also made a similar comparison  with the clover fermion calculation~\cite{Yoon:2016dij} for  $g_A^3$.
 The pion masses and lattice spacings are about the same. We tabulate the relevant parameters in Table~\ref{clover}. 
  \begin{center}
 \begin{table}[htbp]
  \centering
  \caption{The same as in Table~\ref{twisted} for the comparison between clover and overlap fermions. The clover lattice 
  is $32^3 \times 64$ with $m_{\pi} = 312$ MeV, and the overlap lattice is $32^3 \times 64$ with $m_{\pi} = 290$ MeV. The
  inversion time is compared on Titan at OLCF~\cite{RY16}. The total inversion time is in units of node-hrs.} 
 \begin{tabular} {|c|cccccccc|}
  \hline
  Fermion  &  $a$ (fm) &  Config. & Source   & Sink  & $N_{\rm sink}$ & Inversion Time &  LMS & Error   \\
    \hline
   Clover (CI) &    0.081 & 443 & 99 & 2 & 5  & 98(L)+9(H)  & 0 &  1.8\%  \\
    \hline
    Overlap (CI)   & 0.083  & 300 & 1 & 3 & 3 & (1+3$\times$3)$\times$5 & 1/source & 3\% \\
    \hline
 \end{tabular}   \label{clover}
 \end{table}
 \end{center}
To estimate the relative cost, the total inversion times (in node-hours), which includes both the source and sink inversions, are used in Eq.~(\ref{Rcost}) in lieu of $R_{\rm inversion} \times R_{\rm time}$ to obtain $R_{cost} = 1.1$. Again, we see
that the cost of overlap is comparable to that of the clover to reach the same precision.

We also made a comparison with the domain-wall fermion on the same $48^3 \times 64$ lattice at the physical pion mass
for the calculation of $g_A^3$~\cite{Syritsyn:2015nla}. The timing comparison for inversions is based on runs on the pi0 and Ds clusters ar Fermillab~\cite{CS16}. The relevant parameters are listed in Table~\ref{DWF}.

 \begin{center}
 \begin{table}[htbp]
  \centering
  \caption{The same as in Table~\ref{twisted} for the comparison between domain-wall and overlap fermions on the same
  $48^3 \times 96$ lattice at the physical pion mass.  The comparison is made at $t_{sep} = 10$.} 
 \begin{tabular} {|c|cccccc|}
  \hline
  Fermion  &   Config. & Source   & Inversion  & Inversion time  &  LMS & Error   \\
    \hline
   DWF (CI) &   20 & 32(L)+1(H) &  (32+5)$\times$(1+2)  & 1  & 0 &  10\%  \\
    \hline
    Overlap (CI)   & 81 & 5 & 5+8 & $\sim 5.7$ & 8\%/source & 2.2\% \\
    \hline
 \end{tabular}   \label{DWF}
 \end{table}
 \end{center}
 
 The DWF calculation~\cite{Syritsyn:2015nla} used 32 sources with the sloppy inversion and 1 high precision inversion which is 5 times slower. The number of inversions
 in units of the sloppy one is \mbox{32+5 = 37} for the sources and $37 \times 2$ for the sink sequential, where the 2 is for the $u$ and $d$ flavors. Only one mixed polarization is used to get both the polarized and polarized matrix elements. The overlap calculation with the stochastic sandwich algorithm (SSA) used 5 noises for the source and 8 noises
 at the sink at $t_{sep}= 10$. The overlap inversion is $\sim 5.7$ times slower than that of the sloppy DWF. From these
 numbers, we obtain $R_{cost} = 7.4$. This makes the overlap substantially more efficient, mainly due to the 64 grid source
 in one inversion and the fact that the number of inversion in SSA is the sum of those of the source and sink, not their product.



To summarize, we have reported the results of the nucleon matrix elements in the CI on the $48^3 \times 96$ lattice at physical
pion mass with the overlap fermion. With three time separations between the source and the sink, we obtain $g_A^3 = 1.18(4)$.
We also report a global fit including several lattices to take into account the finite volume and finite lattice spacing corrections which gives $\langle x\rangle_{u-d} = 0.170(14)$ in the $\overline{MS}$ scheme at $\mu = 2$ GeV which is consistent with experiment.

We have made a cost comparison of the overlap fermion with several fermion actions in calculating the nucleon matrix elements. Despite a longer inversion time for the overlap fermion, various correlator improvements, such as the stochastic sandwich algorithm with the noise grid source and sink with low-mode substitution for the CI and low-mode average for the quark loop in the DI, have made the overlap as efficient as the twisted mass and clover fermions for the present nucleon matrix element calculation and it is more efficient than the DWF. When the multi-mass feature is employed to include the partially quenched data, the overlap fermion can be more efficient than the single mass comparison made here.


\begin{thebibliography}{99}\bibitem{Li:2010pw} 
  A.~Li {\it et al.} [$\chi$QCD Collaboration],
  Phys.\ Rev.\ D {\bf 82}, 114501 (2010), [arXiv:1005.5424 [hep-lat]].
 %
 \bibitem{Gong:2013vja} 
  M.~Gong {\it et al.} [XQCD Collaboration],
  Phys.\ Rev.\ D {\bf 88}, 014503 (2013),
  [arXiv:1304.1194 [hep-ph]].  
%
\bibitem{Gong:2015iir} 
  M.~Gong, Y.~B.~Yang, A.~Alexandru, T.~Draper and K.~F.~Liu,
  arXiv:1511.03671 [hep-ph].
%
\bibitem{Yang:2015uis} 
  Y.~B.~Yang, A.~Alexandru, T.~Draper, J.~Liang and K.~F.~Liu,
  Phys.\ Rev.\ D {\bf 94}, no. 5, 054503 (2016),
  arXiv:1511.09089 [hep-lat].  
 %
 \bibitem{Blum:2014tka} 
  T.~Blum {\it et al.} [RBC and UKQCD Collaborations],
  Phys.\ Rev.\ D {\bf 93}, no. 7, 074505 (2016),
  [arXiv:1411.7017 [hep-lat]].
 %
 \bibitem{Yang:2015zja} 
  Y.~B.~Yang, A.~Alexandru, T.~Draper, M.~Gong and K.~F.~Liu,
  Phys.\ Rev.\ D {\bf 93}, no. 3, 034503 (2016),
  [arXiv:1509.04616 [hep-lat]].  
 %
 \bibitem{Abdel-Rehim:2015owa} 
  A.~Abdel-Rehim {\it et al.},
  Phys.\ Rev.\ D {\bf 92}, no. 11, 114513 (2015)
  [arXiv:1507.04936 [hep-lat]].  
%
\bibitem{Liang:2016fgy} 
  J.~Liang, Y.~B.~Yang, K.~F.~Liu, A.~Alexandru, T.~Draper and R.~S.~Sufian,
  arXiv:1612.04388 [hep-lat].  
%
\bibitem{Ama91}  
 New Muon Collaboration, P. Amaudruz et al. , Phys. Rev. Lett. 66, 2712 (1991); 
  M. Arneodo et al., Phys. Rev. D 50, R1 (1994).
%
\bibitem{KH15}
Kyriakos Hadjiyiannakou, private communication.

\bibitem{Abdel-Rehim:2016won} 
  A.~Abdel-Rehim, {\it et al.}
  Phys.\ Rev.\ Lett.\  {\bf 116}, no. 25, 252001 (2016),
  [arXiv:1601.01624 [hep-lat]].
  %
\bibitem{Yoon:2016dij} 
  B.~Yoon {\it et al.},
  Phys.\ Rev.\ D {\bf 93}, no. 11, 114506 (2016),
  [arXiv:1602.07737 [hep-lat]].
%
\bibitem{RY16}
R. Gupta and B. Yoon, private communication.
 %
\bibitem{Syritsyn:2015nla} 
  S.~Syritsyn,
  J.\ Phys.\ Conf.\ Ser.\  {\bf 640}, no. 1, 012054 (2015).
  doi:10.1088/1742-6596/640/1/012054 
%
\bibitem{CS16}
C. Lehner, S. Syritsyn, and T. Izubuchi, private communication. 



  \end{thebibliography}
%

  \end{document}